\documentclass[12pt,notitlepage]{article}
\usepackage{nopageno}
\usepackage[top=1in,left=1.25in,right=1.25in,bottom=1in]{geometry}

\usepackage{hyperref}

\usepackage{array,booktabs,multirow,makecell}

\usepackage{amssymb}
\usepackage[T1]{fontenc}
\usepackage{amsmath}

\usepackage{amsthm}

\newcommand{\myparagraph}[1]{\smallskip\noindent {\bf #1.}}

\newcommand{\Str}{\operatorname{\sf S}}
\newcommand{\rank}{\operatorname{\sf rank}}
\newcommand{\select}{\operatorname{\sf select}}
\newcommand{\access}{\operatorname{\sf access}}

\newtheorem{theorem}{Theorem}[section]

\usepackage[affil-it]{authblk} 
\usepackage{float,color,subfigure,graphicx}
\date{}

  \title{Improved Parallel Construction of Wavelet Trees and Rank/Select Structures}
\author{Julian Shun\authorcr MIT CSAIL \authorcr jshun@mit.edu}%

\begin{document}

\maketitle

\begin{abstract}
Existing parallel algorithms for wavelet tree construction have a work
complexity of $O(n\log\sigma)$.  This paper presents parallel
algorithms for the problem with improved work complexity. Our first
algorithm is based on parallel integer sorting and has either
$O(n\log\log n\lceil\log\sigma/\sqrt{\log n\log\log n}\rceil)$ work and
polylogarithmic depth, or $O(n\lceil\log\sigma/\sqrt{\log n}\rceil)$
work and sub-linear depth. We also describe another algorithm that
has $O(n\lceil\log\sigma/\sqrt{\log n} \rceil)$ work and $O(\sigma+\log n)$
depth.
We then show how to use similar ideas to construct variants of wavelet
trees (arbitrary-shaped binary trees and multiary trees) as well as
wavelet matrices in parallel with lower work complexity than prior
algorithms.  Finally, we show that the rank and select structures on
binary sequences and multiary sequences, which are stored on wavelet
tree nodes, can be constructed in parallel with improved work
bounds, matching those of the best existing sequential algorithms for constructing rank and select structures. 
\end{abstract}

\section{Introduction}
The \emph{wavelet tree} is a space-efficient data structure that
supports access, rank, and select queries on a sequence in time
logarithmic in the alphabet size. It was introduced by Grossi et
al.~\cite{Grossi2003}, who used it to design a compressed suffix
array. Wavelet trees have many other
applications~\cite{Navarro2012,Makris12,Navarro2016book,MBCT2015}---for example, they can be
used to obtain compressed representations of sequences, permutations,
grids, graphs, and self-indexes based on the Burrows-Wheeler
transform, and can also be used for two-dimensional
range queries~\cite{MakinenN07}.

The standard sequential method for constructing a wavelet tree on a
sequence of length $n$ with alphabet size $\sigma$ takes
$O(n\log\sigma)$ work.\footnote{We use $\log x$ to mean the base 2
  logarithm of $x$ unless specified otherwise.}   Recently, faster sequential algorithms
with $O(n \lceil\log\sigma/\sqrt{\log n} \rceil)$ work have been
described~\cite{Munro15,BabenkoGKS14} and implemented by
Kaneta~\cite{Kaneta2018}. As for prior parallel algorithms,
Fuentes-Sepulveda et al.~\cite{Fuentes14} presented algorithms that
require $O(n\log\sigma)$ work and $O(n)$ depth (parallel
time). Shun~\cite{Shun2015} improved the result by developing faster
parallel algorithms, including one with $O(n\log\sigma)$ work and
$O(\log n\log\sigma)$ depth. Labeit et al.~\cite{Labeit2017} presented a
more space-efficient version of the algorithm from~\cite{Shun2015}
that achieves the same bounds, as well as a modification of the
algorithm from~\cite{Fuentes14} that has more parallelism.  Later,
Fuentes-Sepulveda et al.~\cite{Fuentes2016} presented a similar
modification to their previous algorithm~\cite{Fuentes14}. Recently,
Fischer et al.~\cite{Fischer2018} presented parallel wavelet tree
construction algorithms that are fast in practice.

While parallel algorithms exist for wavelet tree construction, their
work complexities are higher than those of the best sequential algorithms,
which take $O(n \lceil\log\sigma/\sqrt{\log n} \rceil)$
work~\cite{Munro15,BabenkoGKS14}.  This paper presents parallel
algorithms for wavelet tree construction with improved work
complexities. Our first algorithm is a parallelization of the algorithm
in~\cite{BabenkoGKS14} and uses parallel integer sorting.  Depending
on the parallel integer sorting subroutine used, our algorithm takes
either $O(n\log\log n \lceil\log\sigma/\sqrt{\log n\log\log n} \rceil)$ work and
$O(\log n\log\sigma)$ depth or $O((n/\epsilon) \lceil
\log\sigma/\sqrt{\log n} \rceil))$ work and 
$O((n^\epsilon/\epsilon)\lceil\log\sigma/\sqrt{\log n}\rceil)$
depth for
a constant $0<\epsilon<1$. This results in either a polylogarithmic-depth
algorithm with improved work complexity, or a sub-linear depth algorithm whose work matches that of the best sequential algorithm.
Our second algorithm is based on a simple
domain-decomposition approach as used in~\cite{Fuentes2016,Labeit2017},
and takes $O(\sigma K + n\lceil\log\sigma/\sqrt{\log n}\rceil)$ work
and $O((n/K)\lceil\log\sigma/\sqrt{\log n}\rceil + \log K)$ depth for
any integer $K \ge 1$. Setting
$K=\Theta((n/\sigma)\lceil\log\sigma/\sqrt{\log n}\rceil)$ gives an
algorithm with
$O(n\lceil\log\sigma/\sqrt{\log n}\rceil)$ work and
 $O(\sigma+\log n)$ depth. This algorithm therefore has high parallelism for small alphabet sizes.
We can improve the depth by combining the domain-decomposition approach with our algorithm based on integer sorting, which gives us an algorithm with $O((n/\epsilon)\lceil\log\sigma/\sqrt{\log n}\rceil)$ work and $O((\sigma^\epsilon/\epsilon)\lceil\log\sigma/\sqrt{\log n}\rceil+\log n)$ depth.

Using similar ideas we also obtain improved algorithms for
constructing variants of the standard wavelet tree, such as
arbitrary-shaped binary wavelet trees~\cite{Foschini2006}, multiary
trees~\cite{Ferragina2007}, and wavelet matrices~\cite{Claude12}.  Wavelet tree
nodes store rank and select structures, and so to achieve the improved
work bounds, we show how to construct in parallel the rank and select
structures of binary and multiary sequences work-efficiently.
For binary sequences of length $n$ we show how to construct the
structures in $O(n/\log n)$ work and $O(\log n)$ depth (the sequence
lengths across all wavelet tree nodes sum to $O(n\log \sigma)$, so
this contributes a total of $O(n\log\sigma/\log n)$ work, which is within the desired bound).  For
sequences of length $n$ containing characters in $[0,\ldots,\sigma-1]$ for $\sigma=O(\log^{1/3-\delta}n)$ where $\delta>0$, we show how
to construct the structures in $O(n\log\sigma/\log n)$ work and
$O(\log n)$ depth. The work bounds match those of the sequential
algorithms described in~\cite{BabenkoGKS14}.  This is the most
technically involved part of the paper and obtaining these bounds in
parallel requires carefully packing values into words, working on the
compact representations, constructing appropriate lookup tables, and
defining appropriate operators for prefix sum computations. 
Existing and new bounds for the problems studied in this paper are shown in Table~\ref{table:bounds}.

\begin{table*}[!t]
\footnotesize
\centering
  \setlength{\tabcolsep}{1pt}
\def\arraystretch{1.1}
\begin{tabular}{cccc}
\toprule
Data Structure & Algorithm & Work & Depth\\ \midrule
\multirow{6}{*}{Binary Wavelet Tree} & Sequential~\cite{BabenkoGKS14,Munro15} & $O(n\lceil\frac{\log\sigma}{\sqrt{\log n}}\rceil)$ & -- \\
& \cite{Shun2015,Labeit2017} & $O(n\log\sigma)$ & $O(\log n\log\sigma)$ \\
& \cite{Labeit2017,Fuentes2016}$^\dag$ & $O(n\log\sigma)$ & $O(\sigma+\log n)$\\
& \textbf{This paper} & $O(n\log\log n\lceil\frac{\log\sigma}{\sqrt{\log n\log\log n}}\rceil)$ & $O(\log n\log\sigma)$ \\
& \textbf{This paper} & $O((n/\epsilon)\lceil\frac{\log\sigma}{\sqrt{\log n}}\rceil)$ & $O((n^\epsilon/\epsilon)\lceil\frac{\log\sigma}{\sqrt{\log n}}\rceil)$ \\ 
& \textbf{This paper} & $O(n\lceil\frac{\log\sigma}{\sqrt{\log n}}\rceil)$ & $O(\sigma+\log n)$ \\
& \textbf{This paper} & $O((n/\epsilon)\lceil\frac{\log\sigma}{\sqrt{\log n}}\rceil)$ & $O((\sigma^\epsilon/\epsilon)\lceil\frac{\log\sigma}{\sqrt{\log n}}\rceil+\log n)$ \\
\midrule

& Sequential~\cite{BabenkoGKS14,Munro15} & $O(n\lceil\frac{h}{\sqrt{\log n}}\rceil)$ & -- \\
Arbitrary-shaped Binary& \cite{Shun2015} & $O(nh)$ & $O(h\log n)$  \\
Wavelet Tree (height $h$) & \textbf{This paper} & $O(n\log\log n\lceil\frac{h}{\sqrt{\log n\log\log n}}\rceil)$ & $O(h\log n)$ \\
& \textbf{This paper} & $O((n/\epsilon)\lceil\frac{h}{\sqrt{\log n}}\rceil)$ & $O((n^\epsilon/\epsilon)\lceil \frac{h}{\sqrt{\log n}}\rceil)$ \\

\midrule
\multirowcell{4}{Multiary Wavelet Tree \\ (degree $d=O(\log^{1/3-\delta}n)$ \\ for $\delta > 0$)}
& Sequential~\cite{BabenkoGKS14,Munro15} & $O(n\lceil\frac{\log\sigma}{\sqrt{\log n}}\rceil)$ & -- \\
 & \cite{Shun2015} & $O(n\log \sigma)$ & $O(\log n\log \sigma)$ \\
& \textbf{This paper} & $O(n\log\log n \lceil \frac{\log\sigma}{\sqrt{\log n\log\log n}}\rceil)$ & $O(\log n\log\sigma)$ \\
& \textbf{This paper} & $O((n/\epsilon)\lceil\log\sigma/\sqrt{\log n}\rceil)$ & $O((n^\epsilon/\epsilon)\lceil \frac{\log\sigma}{\sqrt{\log n}}\rceil)$ 
\\
\midrule

\multirow{3}{*}{Wavelet Matrix} 
& \cite{Shun2015} & $O(n\log\sigma)$ & $O(\log n\log\sigma)$ \\
& \textbf{This paper} & $O(n\log\log n\lceil\frac{\log\sigma}{\sqrt{\log n\log\log n}}\rceil)$ & $O(\log n\log\sigma)$ \\
& \textbf{This paper} & $O((n/\epsilon)\lceil\frac{\log\sigma}{\sqrt{\log n}}\rceil)$ & 
$O((n^\epsilon/\epsilon)\lceil \frac{\log\sigma}{\sqrt{\log n}}\rceil)$ \\
\midrule 

\multirow{3}{*}{Binary Rank and Select}
& Sequential~\cite{BabenkoGKS14,Munro15} &  $O(\frac{n}{\log n})$ & -- \\ 
& \cite{Shun2015} & $O(n)$ & $O(\log n)$ \\
& \textbf{This paper} & $O(\frac{n}{\log n})$ & $O(\log n)$ \\
\midrule
\multirowcell{3}{Generalized Rank and Select \\ (degree $d=O(\log^{1/3-\delta}n)$ \\ for $\delta>0$)} & Sequential~\cite{BabenkoGKS14} &  $O(\frac{n\log\sigma}{\log n})$ & -- \\ 
 & \cite{Shun2015} & $O(n)$ & $O(\log n)$ \\
 & \textbf{This paper} & $O(\frac{n\log\sigma}{\log n})$ & $O(\log n)$ \\

\bottomrule
\end{tabular}
\caption{New and existing work and depth bounds for constructing data structures. We omit the depth term for the sequential algorithms. $^\dag$A parameter in the algorithm was chosen to give the minumum depth while maintaining $O(n\log\sigma)$ work for any $\sigma$.}
\label{table:bounds}
\end{table*}

\section{Preliminaries}\label{sec:prelims}
We analyze algorithms in the \emph{work-depth model}, where the
\emph{work} $W$ is the number of operations required (equivalent to
the standard sequential time complexity) and the \emph{depth (parallel time)} $D$ is
the length of the longest critical path in the
computation~\cite{Vishkin10}.  The parallelism (maximum possible speedup) of an algorithm is equal
to $W/D$.  With $p$ available processors, using Brent's scheduling
theorem~\cite{Brent1974} we can bound the running time by $W/p + D$.
We say that a parallel algorithm is
\emph{work-efficient} if its asymptotic work complexity matches that
of the best sequential algorithm.
As in the standard word RAM model, we assume that
$\Theta(\log n)$ bits fit in a word, and reading or writing a word requires
unit work.

\begin{figure}[!t]
\centering
\includegraphics[width=\linewidth]{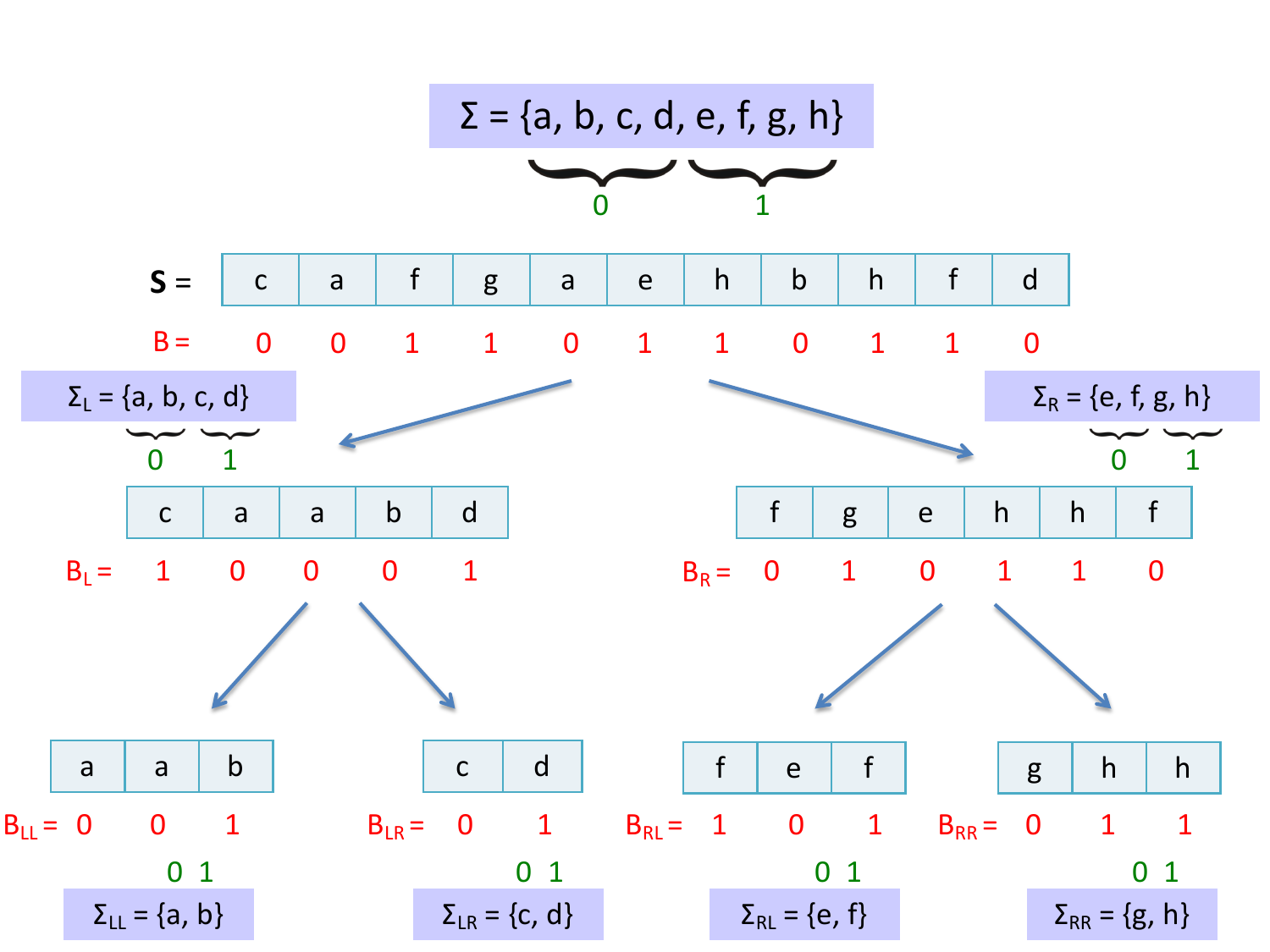}
\caption{Example of a wavelet tree on the
  sequence $\Str=\texttt{cafgaehbhfd}$ and where $\sigma=8$ and the alphabet is
  $\Sigma=\{\texttt{a},\texttt{b},\texttt{c},\texttt{d},\texttt{e},\texttt{f},\texttt{g},\texttt{h}\}$. The root contains the entire
  sequence as well as a bitstring indicating whether the
  symbol comes from the left or the right half of $\Sigma$. The
  two children $L$ and $R$ of the root contain the characters in the
  left and right half of $\Sigma$, respectively. The bitstring for the
  left (right) child $B_L$ ($B_R$) is constructed by checking if the character comes
  from the left half or the right half of $\Sigma_L$ ($\Sigma_R$). The
  leaves each represent an alphabet of size $2$.}
\label{fig:wavelet}
\end{figure}

A sequence of symbols will be denoted by $\Str$, its length by $n$,
and its alphabet size by $\sigma$.
For a sequence $\Str$, $\access(\Str,i)$ returns the symbol at position
$i$ of $\Str$, $\rank_c(\Str,i)$ returns the number of times $c$ appears in
$\Str$ from positions $0$ to $i$, and $\select_c(S,i)$ returns the
position of the $i$'th occurrence of $c$ in $\Str$.  A \emph{wavelet
  tree} is a data structure supporting access, rank, and select
operations on a sequence in $O(\log \sigma)$
work~\cite{Grossi2003}. A standard wavelet
tree is a balanced binary tree where each node represents a range of
the symbols in the alphabet using a bitstring (binary sequence).  We
assume that $\sigma \le n$, and that the alphabet is
$[0,\ldots,\sigma-1]$, as the symbols can be mapped to a contiguous
range otherwise.  The structure of the wavelet tree is defined
recursively as follows: The root represents the symbols
$[0,\ldots,2^{\lceil \log \sigma \rceil}-1]$.  A node $v$ which
represents the symbols $[a,\ldots,b]$ stores a bitstring which has a $0$
in position $i$ if the $i$'th symbol in the range $[a,\ldots,b]$ is in
$[a,\ldots,(a+b+1)/2-1]$ and $1$ otherwise. It will have a left child
that represents the symbols $[a,\ldots,(a+b+1)/2-1]$ and a right child
that represents the symbols $[(a+b+1)/2,\ldots,b]$. The recursion
stops when the range is of size $2$ or less or if a node has no symbols
to represent. 
An example of a wavelet tree is shown in
Figure~\ref{fig:wavelet}. 
We point out that the
original wavelet tree description in~\cite{Grossi2003} uses a root
whose range is not necessarily a power of $2$, but the definition here
gives the same asymptotic query times and leads to a simpler
description of our construction algorithms.

Each node in the wavelet tree stores a \emph{succinct} rank/select
structure on its bitstring (whose size is sub-linear in the bitstring
length) to allow for constant-work rank and select queries. The
bitstrings and the rank/select structures together take $n\lceil \log
\sigma \rceil + o(n\log\sigma)$ bits of space. The wavelet tree
topology requires $O(\sigma\log n)$ bits to store pointers, but this
can be removed by modifying how the queries are
performed~\cite{MakinenN07,ClaudeN08}.

Our algorithms use prefix sum as a parallel
primitive~\cite{Vishkin10}. \emph{Prefix sum} takes as input an
array $X$ of length $n$, an associative binary operator $\oplus$, and
an identity element $\bot$ such that $\bot \oplus x = x$ for any $x$,
and returns the array $(\bot,\bot \oplus X[0], \bot \oplus X[0] \oplus
X[1], \ldots,\bot \oplus X[0] \oplus X[1] \oplus \ldots \oplus
X[n-2])$, as well as the overall sum $\bot \oplus X[0]\oplus X[1]
\oplus \ldots \oplus X[n-1]$. Assuming that $\oplus$
takes constant work, prefix sum can be implemented in $O(n)$ work and
$O(\log n)$ depth~\cite{Vishkin10}.  Unless specified otherwise, we will
use $\oplus$ to be the addition operator on integers.

\begin{figure}[!t]
\centering
\includegraphics[width=\linewidth]{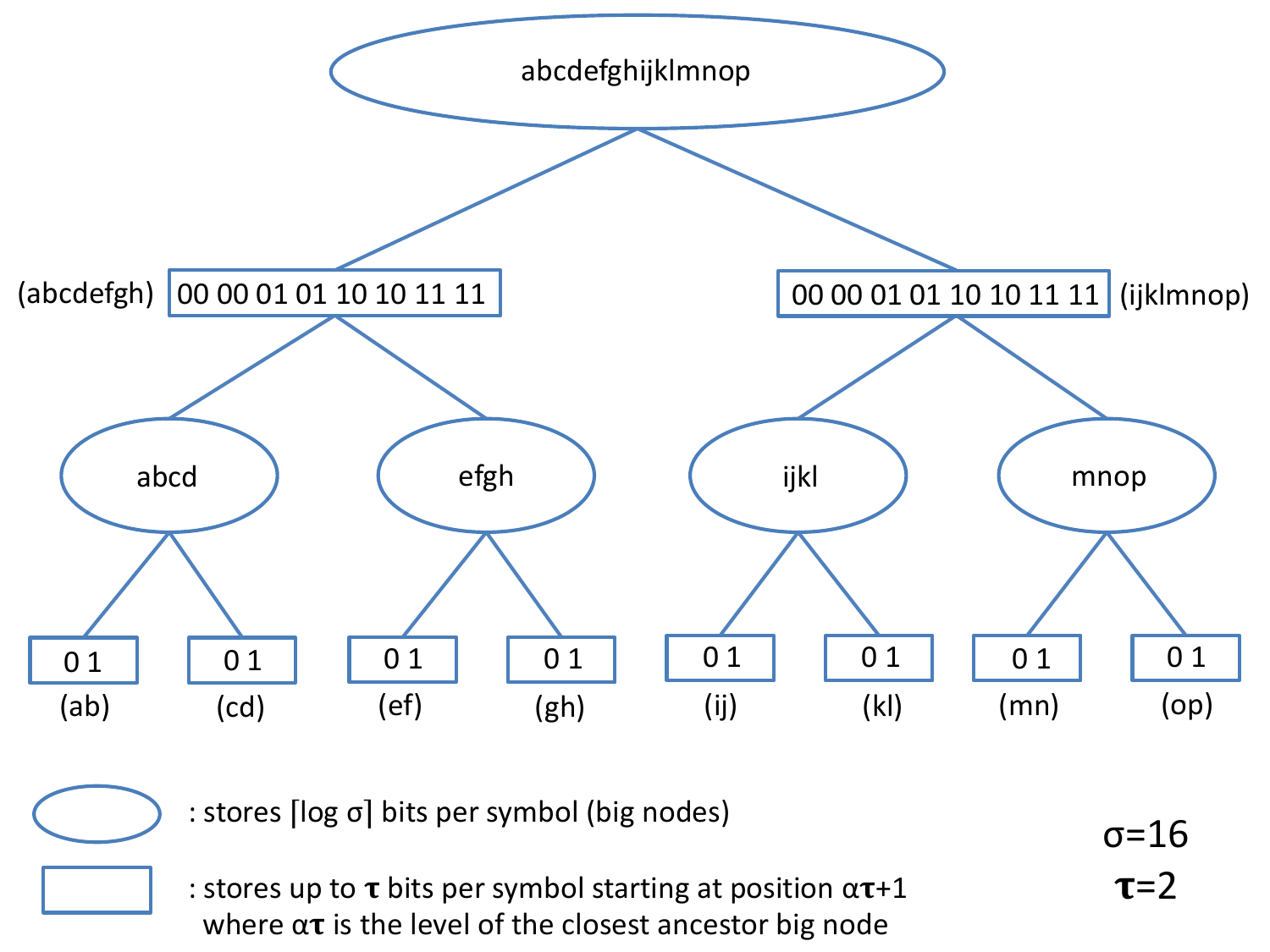}
\caption{Example of the two types of nodes in the algorithm of Babenko et al.~\cite{BabenkoGKS14} for $\tau=2$ and $\sigma=16$. The alphabet is
  $\Sigma=\{\texttt{a},\texttt{b},\texttt{c},\texttt{d},\texttt{e},\texttt{f},\texttt{g},\texttt{h},\texttt{i},\texttt{j},\texttt{k},\texttt{l},\texttt{m},\texttt{n},\texttt{o},\texttt{p}\}$ and sequence is $\Str=\texttt{abcdefghijklmnop}$.
  The ovals correspond to big nodes, which store $\lceil\log\sigma\rceil=4$ bits per symbol. The rectangles correspond to the other nodes, which store up to $\tau=2$ bits per symbol in short lists (the symbols they represent are shown in parentheses next to the node).
 The bitstrings per node are not shown.}
\label{fig:alg1-figure}
\end{figure}

\section{Review of the $O(n\lceil\log\sigma/\sqrt{\log n}\rceil)$ Work Sequential Algorithm}\label{sec:review}
We first review how the $O(n\lceil\log\sigma/\sqrt{\log n}\rceil)$ work sequential
wavelet tree construction algorithm from~\cite{BabenkoGKS14} works, as
we will be parallelizing this algorithm.  A similar sequential algorithm was
independently described in~\cite{Munro15}. Figure~\ref{fig:alg1-figure} illustrates the two types of nodes in the algorithm.
The basic data structure used is a \emph{packed list}, which
stores $N$ $b$-bit integers using $\lceil Nb/\log n \rceil$ words. It supports
appending a length $N$ list in $O(\lceil Nb/\log n \rceil)$ work and splitting a
list into smaller lists of at most length $k$ in $O(\lceil Nb/\log n \rceil + N/k)$
work. These operations can be implemented using bit-shifts and copying.
In this wavelet tree algorithm, a \emph{big node} is defined to be a node at a level that is a
multiple of $\tau$, where $\tau$ is a parameter to be chosen. A big
node stores the symbols that it represents in $\Str$, using
$\lceil\log\sigma\rceil$ bits per symbol as in the standard representation. Big nodes
can be computed recursively as follows. The root is a big node storing $\Str$. Assume
that the sub-sequences for the big nodes at level $\alpha\tau$ are
already computed. Then to compute the symbols in the big nodes at
level $(\alpha+1)\tau$, the big nodes at level $\alpha\tau$ look
at the $\tau$ bits starting at position $\alpha\tau$ in the binary
representation of each symbol to determine which of its descendant
big nodes at level $(\alpha+1)\tau$ to place the symbol at (there are
$2^\tau$ such descendants). Therefore, computation for big nodes requires
$O(n \lceil\log\sigma/\tau \rceil)$ work overall.

Nodes at all other
levels of the tree only need to store at most $\tau$ bits per symbol (the $\tau$ bits
starting at position $\alpha\tau+1$, where $\alpha\tau$ is the level of
its nearest big node ancestor) because there are only $\tau$ levels
between two big node levels.
These nodes use \emph{short lists} to store $\tau$-bit integers
containing the $\tau$ relevant bits of the symbols they
represent. These are stored as packed lists.  Computing the bitstring
values and short lists is done recursively. The short lists of the
children of a big node can be computed by extracting the relevant bits
from the symbols of the big nodes in $O(n \lceil\log\sigma/\tau \rceil)$ work across
all big nodes. Given a short list of a node, computing its own bitstring
values and the short lists of its children is done via table lookup.
For all packed lists $L$ of at most $\log n/(2\tau)$ $\tau$-bit integers, the
bitstring value, and the packed lists $L_0$ and $L_1$ consisting of the
symbols of $L$ whose $t$'th most significant bit is $0$ or $1$,
respectively, are pre-computed for all $t\in [0,\tau-1]$. Pre-computing
this table involves evaluating all $O(2^{\log n/2})$ $\tau$-bit integer sequences
of length at most $\log n/(2\tau)$ for each value of $t$. This can be done in $O(n)$ work.
Each node splits its short list
into blocks of length at most $\log n/(2\tau)$, performs table
lookups for each block, and then appends the resulting bitstring values
together, $L_0$'s together, and $L_1$'s together. The bitstring values are
stored in the bitstring associated with the current node, and $L_0$ and
$L_1$ are passed to its children.  For a node with a short list of
length $N$, the total work required is $O(N\tau/\log n)$ as the
splitting and merging can be done in $O(N\tau/\log n)$ work overall
and table lookups in constant work per block.  The sum of the lengths of all
short lists is $O(n\log\sigma)$, and so the total work required for this
computation is $O(n \lceil\log\sigma(\tau/\log n) \rceil)$.

The overall work is $O(n \lceil\log\sigma/\tau \rceil + n
\lceil\log\sigma(\tau/\log n) \rceil)$ and choosing $\tau = \sqrt{\log
  n}$ minimizes the work, giving a bound of $O(n
\lceil\log\sigma/\sqrt{\log n} \rceil)$. By constructing the tree
level-by-level (i.e., interleaving big node computation with levels in
between big nodes), at any time the algorithm only has to store the symbols for
the big nodes at one level and short lists at one level, and so the peak
space usage of the algorithm is $O(n\log\sigma)$ bits.

\section{Parallel Wavelet Tree Algorithms}\label{sec:wt}
This section first describes how to parallelize the algorithm
of Babenko et al.~\cite{BabenkoGKS14}, which we reviewed in Section~\ref{sec:review}. Then we present a simple domain-decomposition
based parallel construction algorithm that is work-efficient and whose
parallelism depends linearly on $\sigma$, and so has low depth for small
alphabets.

\subsection{Parallelizing the algorithm of Babenko et al.~\cite{BabenkoGKS14}}\label{sec:firstalg}
The nodes in our parallel algorithm are classified the same way as in the sequential algorithm (see Figure~\ref{fig:alg1-figure}).
The sub-sequences for the big nodes can be computed level-by-level
using parallel integer sorting. In particular, given the correct
sub-sequence $\Str'$ for a big node at level $\alpha\tau$, we
compute the sub-sequences for its big node descendants at level
$(\alpha+1)\tau$ by performing an integer sort on $\Str'$,
where the key for the sort is the value of the (up to) $\tau$
bits starting from the $\alpha\tau$'th highest bit of the
symbol.

The parallel integer sort that we use is required to be stable since we
need to keep the relative ordering among the characters in each
descendant node. Unfortunately the only known method for stable parallel
integer sorting in linear work and polylogarithmic depth~\cite{RR89}
requires the range of the keys of the values being sorted to be
polylogarithmic, which does not hold for the value of $\tau$ that we will choose.  
Instead we can
either use an algorithm that is not work-efficient, requiring
$O(n\log\log n)$ work and $O(\log n)$
depth~\cite{Raman1990,Bhatt1991},\footnote{These algorithms either use
  randomization~\cite{Raman1990} or require super-linear space~\cite{Bhatt1991}.} or use a work-efficient algorithm
with $O(n/\epsilon)$ work and
$O(n^\epsilon/\epsilon)$ depth for
a constant $0<\epsilon<1$~\cite{Vishkin10}.  This gives an overall complexity for constructing big nodes of
either (a) $O(n\log\log n\lceil \log\sigma/\tau \rceil)$ work and $O(\log n\lceil\log\sigma/\tau\rceil)$ depth or (b)
$O((n/\epsilon)\lceil\log\sigma/\tau\rceil)$ work and
$O((n^\epsilon/\epsilon)\lceil\log\sigma/\tau\rceil)$
depth for constructing the big nodes.

The lookup table for computing short lists can be pre-computed by evaluating all
$O(2^{\log n/2})$ $\tau$-bit integer sequences of length at most $\log n/(2\tau)$ for each $t\in [0,\tau-1]$ in parallel, and storing the answer for each in a unique
location. For example, this can be done using a three-level table, with the first level indexed by sequence length, second level by $t$, and third level by the value of the sequence as an integer. 
The result for each sequence and value of $t$ is evaluated sequentially. Overall, this requires
$O(\log n)$ depth and $o(n)$ work.

Computing short lists for children of a big node can be done in linear
work and $O(\log n)$ depth by extracting the relevant bits from the
symbols in the big node, performing prefix sums to get the appropriate
offsets, and copying the $\tau$ bits of a symbol into the appropriate location in an
array of the appropriate child in parallel.  Groups of $\tau$-bit integers that together form a word are then packed together and
copied into one entry of the short list for the corresponding child in
parallel.  The bitstrings of the children of a big node can be computed in linear work and $O(\log n)$ depth simply by extracting the relevant bit from the symbols and packing them together. 
Computing short lists of other nodes requires merging and
splitting packed lists. For each short list, we split it
into chunks containing at most $\log n/(2\tau)$ $\tau$-bit integers by copying the
relevant bits of each chunk into its own word in constant depth. The
algorithm performs a table lookup for each chunk to obtain the parts
of the packed lists $L_0$ and $L_1$ that the chunk contributes to as
well as the part of the bitstring associated with the chunk.  All table
lookups are done in parallel in constant depth.  We then merge
together the results to form each of $L_0$, $L_1$, and the bitstring for
the node.  To merge the results of one of the lists together, we
compute the length (in bits) of the result associated with each chunk,
perform a prefix sum to determine the total length (in bits) and also
the offset for each result in a new array, and allocate a new array of
the desired length.  We then identify the groups of chunks that will
copy into the same word, again using prefix sums (some chunks will copy into two words, but this only increases the work by a constant factor). Then, in parallel,
all groups merge their chunks sequentially using the packed list operations described in Section~\ref{sec:review} and then copy their word
into the new array at the appropriate offset.
There are a total of $\lceil 2N\tau/\log n\rceil$
chunks if the short list contains $N$ integers, each of which
generates a partial result for $L_0$, $L_1$, and the bitstring, and so the
prefix sum and copying takes $O(N\tau/\log n)$ work and $O(\log n)$
depth (there is a constant-factor overhead due to some chunks not being full, however the complexity is not affected).
The overall work for computing the
short lists is $O(n \lceil\log\sigma(\tau/\log n) \rceil)$ as in the sequential
algorithm.  The depth is $O(\log n\log\sigma)$ as there are
$\log\sigma$ levels, each requiring $O(\log n)$ depth.

To minimize the overall work we set $\tau=\sqrt{\log n\log\log n}$
when using the $O(n\log\log n)$ work integer sort and $\tau=\sqrt{\log
  n}$ when using the $O(n/\epsilon)$ work integer sort.  Assuming that
constructing the rank and select data structures per node can be done
in the same bounds, which we describe in Section~\ref{sec:rankselect},
we obtain the following theorem:
\begin{theorem}\label{thm:1}
Wavelet tree construction can be performed in 
$O(n\log\log n\lceil\log\sigma/ \allowbreak\sqrt{\log n\log\log n}\rceil)$ work and $O(\log n\log\sigma)$ depth (using randomization or super-linear space)
or $O((n/\epsilon)\lceil\log\sigma/\sqrt{\log n}\rceil)$ work and
$O((n^\epsilon/\epsilon)\lceil\log\sigma/\sqrt{\log n}\rceil)$
depth
for a constant $0<\epsilon<1$.
\end{theorem}

Note that both parallel algorithms described above improve upon the
$O(n\log\sigma)$ work complexity of the algorithms described
in~\cite{Shun2015,Labeit2017}. Our algorithm either has polylogarithmic
depth but does not achieve the $O(n\lceil\log\sigma/\sqrt{\log n}\rceil)$ work
bound of the best sequential algorithm, or is
work-efficient with sub-linear (but not polylogarithmic) depth. However,
as long as the number of processors is sub-linear, the second
algorithm can make full use of all of the available processors (recall
Brent's scheduling theorem and the definition of parallelism from
Section~\ref{sec:prelims}). Improving parallel integer sorting
algorithms would immediately improve the complexity of the wavelet
tree algorithms.

We now analyze the working space of the algorithm. We also compute the tree level-by-level as in the sequential algorithm. We require $O(n\log n)$ bits of working space for the integer sort (assuming that we use~\cite{Vishkin10} or~\cite{Raman1990}). The prefix sums and packing operations also require $O(n\log n)$ bits of working space. Finally, the lookup table contains $o(n)$ entries and therefore uses $o(n\log n)$ bits. Overall, our algorithm requires $O(n\log n)$ bits of working space.

\subsection{Domain-decomposition approach} 
\begin{figure}[!t]
\centering
\includegraphics[width=\linewidth]{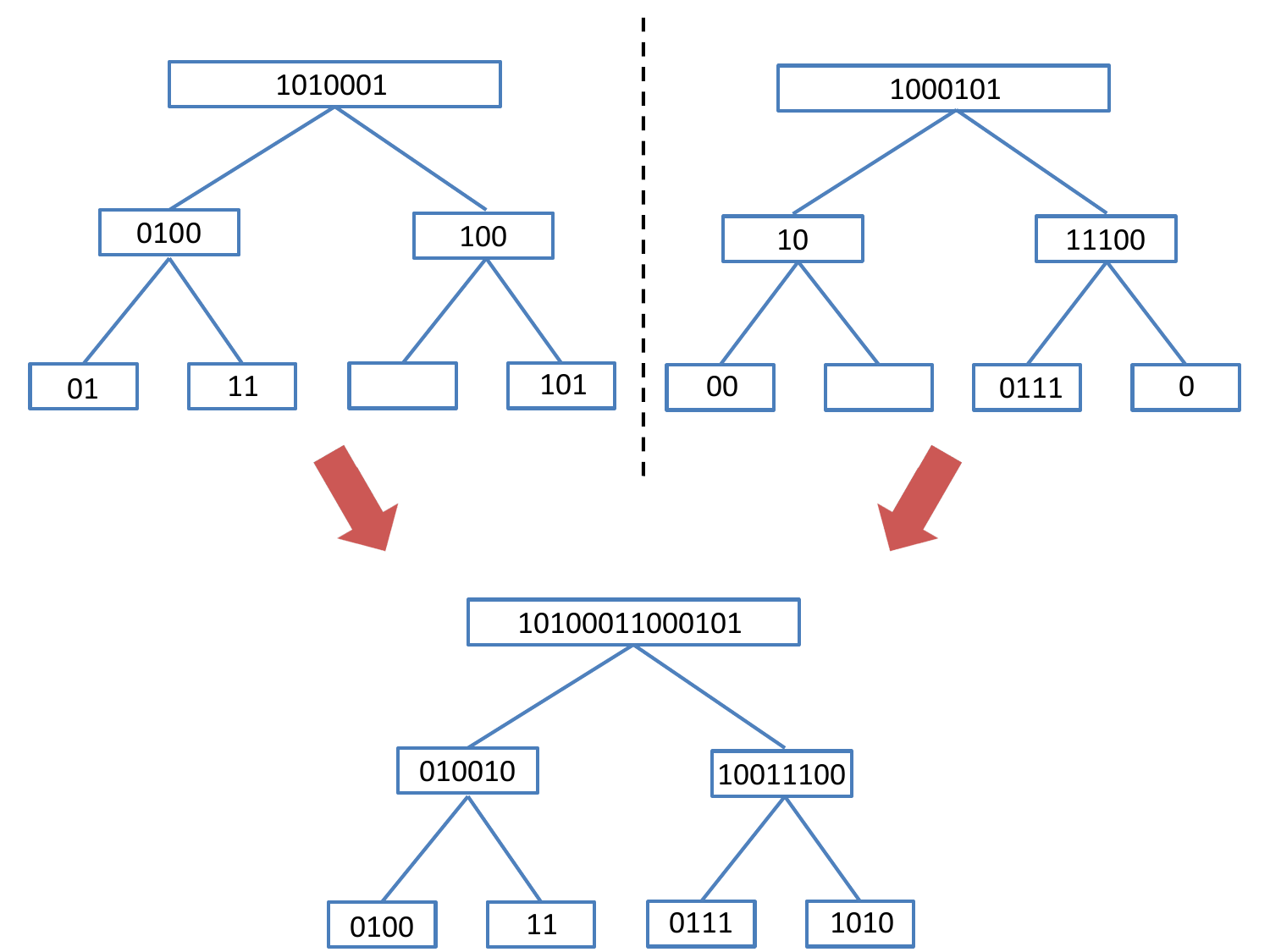}
\caption{An illustration of the domain-decomposition approach where $K=2$. The sequence is split in half, and the wavelet trees, including bitstrings per node, are generated independently for each half (top). The bitstrings are then merged together for each node to obtain the final wavelet tree (bottom).  }
\label{fig:alg2-figure}
\end{figure}

Another way to construct the wavelet tree in parallel is to use a
domain-decomposition approach as done in~\cite{Fuentes2016,Labeit2017}. For a parameter $K$, this approach first splits
the input sequence into $K$ equal-sized sub-sequences, constructs the wavelet tree (without rank/select structures) across
all sub-sequences in parallel using a sequential algorithm for each, and then
merges the bitstrings on the nodes of the $K$ trees
together. An illustration of the domain-decomposition approach is shown in Figure~\ref{fig:alg2-figure}. Constructing the tree for each sub-sequence can be done by using an
$O(n\lceil\log\sigma/\sqrt{\log n}\rceil)$ work sequential
algorithm~\cite{Munro15,BabenkoGKS14} in a black-box fashion (where
the alphabet size for each sub-sequence is treated as the same as the
alphabet size of the entire sequence). The overall work for this step
is $O(n \lceil\log\sigma/\sqrt{\log n}\rceil)$ and the depth is
$O((n/K)\lceil\log\sigma/\sqrt{\log n}\rceil)$.

To merge together the bitstrings, we first form the wavelet tree structure (without bitstrings on nodes), which takes $O(\sigma)$ work and $O(1)$
depth. Following the idea described in~\cite{Fuentes2016,Labeit2017}, for each
node in the final tree structure, we then perform a prefix sum across
the lengths of the bitstrings on the corresponding nodes in the $K$
sub-problems (the length is $0$ if the node does not exist) taking $O(K)$ work and $O(\log K)$ depth. This
gives the length of the bitstring on the node in the final tree as
well as an appropriate offset into the bitstring for each
sub-problem. Then each sub-sequence copies its bitstring into the
bitstring of the node in the final tree in parallel at word granularity. The words
where multiple sub-sequences can copy into in parallel are marked
beforehand to avoid conflicts and handled specially (these ``boundary'' words can be identified by looking at the offsets of the $O(\sigma K)$ nodes, and there can be at most $O(\sigma K)$ of them). 
Summed over all nodes in the final tree, the prefix sums take
$O(\sigma K)$ work and $O(\log K)$ depth (the $\sigma$ different
prefix sums can be done independently in parallel).  
Excluding the special words, the copying
takes $O(n\log\sigma/\log n)$ work and $O(1)$
depth in total (the $\log n$ in the denominator of the work is
because we are copying at word granularity).  The special words can
all be computed in parallel, taking $O(\sigma K)$ work and $O(\log K)$
depth by concatenating the up to $K$ bitstrings for each special word
in a binary fashion. This gives the following theorem:

\begin{theorem}
A wavelet tree can be constructed in $O(\sigma K +
n \lceil\log\sigma/\sqrt{\log n}\rceil)$ work and
$O((n/K)\lceil\log\sigma/\sqrt{\log n}\rceil +
\log K)$ depth for any integer $K \ge 1$.
\end{theorem}

The domain-decomposition algorithm is work-efficient if $K =
O((n/\sigma)\lceil\log\sigma/\sqrt{\log n}\rceil)$.  Setting
$K=\Theta((n/\sigma)\lceil\log\sigma/\sqrt{\log n}\rceil)$ gives the maximum
parallelism while achieving work-efficiency, and gives a depth of
$O(\sigma+\log n)$. Thus this algorithm has good parallelism for small
$\sigma$, and achieves lower work than the domain-decomposition algorithm in~\cite{Fuentes2016,labeit2015,Labeit2017}.

The space required by the sequential algorithm across all sub-sequences is $O(n\log\sigma)$ bits. The domain-decomposition algorithm also requires $O(\sigma K\log n)$ bits of working space to represent the nodes of the trees of the sub-sequences and for the prefix sums. 
By setting $K=\Theta((n/\sigma)\log\sigma/\log n)$, the space usage does not asymptotically exceed the size of the final output of $O(n\log\sigma)$ bits, the work is $O(n\lceil \log\sigma/\sqrt{\log n}\rceil)$ and the depth is $O(\sigma\log n)$.

We can use a parallel algorithm to solve each of the $K$ sub-problems to improve the depth. In particular, if we plug in the $O((n/\epsilon)\lceil\log\sigma/\sqrt{\log n}\rceil)$ work and $O((n^\epsilon/\epsilon)\lceil\log\sigma/\sqrt{\log n}\rceil)$ depth algorithm from Section~\ref{sec:firstalg} into our domain-decomposition algorithm, we obtain the following theorem.

\begin{theorem}
A wavelet tree can be constructed in $O(\sigma K +
(n/\epsilon) \lceil\log\sigma/\sqrt{\log n}\rceil)$ work and
$O(((n/K)^\epsilon/\epsilon)\lceil\log\sigma/\sqrt{\log n}\rceil +
\log K)$ depth for any integer $1 \le K \le n/\sigma$ and a constant $0<\epsilon<1$.
\end{theorem}

The upper bound on $K$ is due to the fact that integer sort  takes linear work in both the sub-problem size as well as the range of keys being sorted. The range of keys being sorted is $O(\sigma)$, and so we need each sub-problem size to be $\Omega(\sigma)$ to amortize the work to the subproblem size and maintain work-efficiency.
By setting $K=\Theta(n/\sigma)$, we obtain an algorithm with
$O((n/\epsilon)\lceil\log\sigma/\sqrt{\log n}\rceil)$ work and
$O((\sigma^\epsilon/\epsilon)\lceil\log\sigma/\sqrt{\log n}\rceil+\log n)$
depth. The working space is $O(n\log n)$ bits, due to the use of parallel integer sort.

\subsection{Variants}\label{sec:extensions}
This section describes how ideas from our binary wavelet tree construction algorithm from Section~\ref{sec:firstalg} can be used to construct variants of wavelet trees. 

\myparagraph{Arbitrarily-shaped binary trees} Our algorithm from Section~\ref{sec:firstalg} can be extended to binary trees of other shapes
(e.g., Huffman-shaped wavelet trees~\cite{Foschini2006}) if the tree
structure can be computed efficiently and is of height $O(\log n)$. In particular, the algorithm
needs a codeword for each symbol determined by the path from the root
to the node representing the symbol in the tree. The codeword is a
bitstring, where the $i$'th most significant bit is $0$ if the
$(i+1)$'st node in the path is a left child of the $i$'th node in the
path, and is $1$ otherwise. We define $h$ to be the height of the tree.
We assume a
lookup table storing a mapping from codeword to symbol. Since the
codewords are of length $O(\log n)$, we can access the
codeword in constant-work, and construct the lookup table in
$O(\sigma)$ work and $O(\log n)$ depth. (We note that codewords
for a Huffman-shaped wavelet tree can be generated in $O(n)$
work and $O(\sigma+\log n)$ depth~\cite{Edwards2014,Shun2015}.)

To construct the tree, we first convert the symbols to their codewords.
The algorithm proceeds as before, where big nodes are constructed
every $\tau$'th level in the tree by using integer sorting on
$\tau$ bits. Some of the combinations of the bits may not
correspond to a symbol (which can be determined using the lookup table), and no big nodes are generated for those
combinations. The complexity per level is equal to the complexity of
integer sorting, and summing across all $h/\tau$ levels gives
the following bounds for constructing big nodes: (a) $O(n\sqrt{\log\log
n}\lceil h/\sqrt{\log n}\rceil)$ work and $O(\log n\lceil h/\sqrt{\log n\log\log n}\rceil)$
depth (by setting $\tau=\sqrt{\log n\log\log n}$) or (b) $O((n/\epsilon)\lceil h/\sqrt{\log n}\rceil)$ work and
$O((n^\epsilon/\epsilon)\lceil h/\sqrt{\log n}\rceil)$ depth for
$0<\epsilon<1$ (by setting $\tau=\sqrt{\log n}$). The remaining nodes that exist (which again can be checked using the lookup table) are computed using short lists as
before, and the overall work for these nodes is $O(n \lceil
h (\tau/\log n) \rceil)$ and depth is $O(h\log n)$. This gives the
following theorem, whose work bound improves upon the parallel
construction described in~\cite{Shun2015}:

\begin{theorem}\label{thm:2}
Given codewords for the symbols, a binary wavelet tree of height $h=O(\log n)$
can be constructed in $O(n\log\log n \lceil h/\sqrt{\log n\log\log n} \rceil)$ work and
$O(h\log n)$ depth or $O((n/\epsilon) \lceil h/\sqrt{\log n}\rceil)$ work and
$O((n^\epsilon/\epsilon)\lceil h/\sqrt{\log n}\rceil)$ depth
for a constant $0<\epsilon<1$.
\end{theorem}

The working space of the algorithm can be bounded by $O(n\log n)$ bits, as in Section~\ref{sec:firstalg}.

\myparagraph{Multiary wavelet trees} We now describe how to extend the
algorithm to construct multiary wavelet trees~\cite{Ferragina2007} of
degree $d=O(\log^{1/3-\delta} n)$, where $\delta > 0$ and $d$ is a power of two.\footnote{The
  restriction $d=O(\log^{1/3-\delta}n)$ for $\delta > 0$ is due to the rank and select
  structures from~\cite{BabenkoGKS14} that we parallelize.} Each node
now has $d$ children and the sequence that a node stores contains
values in $[0,\ldots,d-1]$ instead of being binary as in the standard
wavelet tree. We describe the algorithm for balanced trees but the
ideas also apply to trees of arbitrary shapes as long as the codewords
are provided as input. Similar to the approach of~\cite{Munro15} we
generate the full binary tree, but only keep sequences for the nodes
at levels $\beta\log d$ in the full binary tree for $\beta =
[0,\ldots,\log\sigma/\log d)$. Each node with a sequence that is kept belongs
  to the multiary wavelet tree, and if it is at level $\beta\log d$ in
  the binary tree, its $d$ children are at level $(\beta+1)\log d$ in
  the binary tree. With an appropriate numbering scheme (i.e., the
  children of node $i$ are stored at locations $2i+1$ and $2i+2$), the
  $d$ children of a node can be identified in $O(d)$ work and $O(1)$
  depth, contributing $O(\sigma)$ work and $O(1)$ depth overall. Each node belonging to the multiary wavelet tree stores a sequence of $\log d$-bit integers, which can be computed by extracting the appropriate $\log d$ bits from its sequence of symbols.
The
  bounds from Theorem~\ref{thm:1} then apply, giving the following
  theorem which improves upon the work of the
  parallel algorithm for multiary wavelet trees from~\cite{Shun2015}.

\begin{theorem}\label{thm:3}
A multiary wavelet tree of degree $d=O(\log^{1/3-\delta}n)$ where $\delta > 0$ and $d$ is a
power of two can be constructed in $O(n\log\log
n\lceil\log\sigma/\sqrt{\log n\log\log n}\rceil)$ work and $O(\log n\log\sigma)$
depth or $O((n/\epsilon)\lceil\log\sigma/\sqrt{\log n}\rceil)$ work
and 
$O((n^\epsilon/\epsilon)$ $\lceil\log\sigma/\sqrt{\log n}\rceil)$
depth for $0<\epsilon<1$.
\end{theorem}

As in Section~\ref{sec:firstalg}, the working space of the algorithm
can be bounded by $O(n\log n)$ bits.  We note that each node of a
multiary wavelet tree requires storing a generalized rank and select
structure on its sequence of $\log d$-bit integers, and we describe
how to construct the structures within the bounds of
Theorem~\ref{thm:3} in Section~\ref{sec:generalized}.

\myparagraph{Wavelet matrix} The wavelet matrix~\cite{Claude12} is a
variant of the wavelet tree where for level $l$, all symbols with a
$0$ as their $l$'th highest bit are represented on the left side of
the level's sequence and all symbols with a $1$ as their $l$'th
highest bit are represented on the right side. The relative ordering
among the symbols from the previous level is preserved.  Each level also contains an integer indicating the number of $0$'s per level. The wavelet
matrix has $O(\log\sigma)$ levels.  An $O(n\log\sigma)$ work,
polylogarithmic depth parallel algorithm for constructing the wavelet
matrix was described in~\cite{Shun2015}. In this section, we describe
how to reduce the work complexity
using similar ideas as described in Section~\ref{sec:firstalg}.

We will process the bits of the symbols in chunks of $\tau$
bits and construct the matrix level-by-level. Every $\tau$'th level is treated specially, similar to the big nodes in Section~\ref{sec:firstalg}. For an integer $\alpha$, to construct the sequence at level $(\alpha+1)\tau$ from level $\alpha\tau$ we perform an integer sort on the sequence at level $\alpha\tau$ using the reverse of the $\tau$ bits starting at the $\alpha\tau$'th position of the symbols. Constructing all special levels takes 
either (a) $O(n\log\log n\lceil \log\sigma/\tau \rceil)$ work and $O(\log n\lceil\log\sigma/\tau\rceil)$ depth or (b)
$O((n/\epsilon)\lceil\log\sigma/\tau\rceil)$ work and
$O((n^\epsilon/\epsilon)\lceil\log\sigma/\tau\rceil)$ depth.

Constructing levels
  $\alpha\tau+1$ to $(\alpha+1)\tau-1$ of the wavelet
  matrix will require only the (at most) $\tau$ bits starting at the $(\alpha\tau+1)$'th position of the symbols. We will create chunks of $\log n/(2\tau)$ $\tau$-bit integers, and use
   the packed list representation as in the
  wavelet tree algorithm.  We use a lookup table storing all possible
  bitstrings of up to length $\log n/2$, which for each chunk and each bit position
  determines which symbols go to the left and which go to the right,
  as well as the bitstring, in $O(1)$ work. The lookup table can be
  computed in $O(\log n)$ depth and $o(n)$ work. Similar to the
  wavelet tree algorithm, each chunk can be split into two parts, the
  first that goes to the left side of the sequence and the second that
  goes to the right. Prefix sums and grouping of chunks are then used
  on the packed lists to create the bitstring for the current level as
  well as the sequence at the next level.
On each level, this takes
  $O(n\lceil\tau/\log n\rceil)$ work and $O(\log n)$ depth. Summing over all
  levels gives $O(n\lceil\log\sigma(\tau/\log n)\rceil)$ work and
  $O(\log n\log\sigma)$ depth. 

  To compute the number of $0$'s in the bitstring for each level, we create a lookup table mapping all possible bitstrings of up to length $\log n/2$ to the number of $0$'s in the bitstring. This can be constructed in $O(\log n)$ depth and $o(n)$ work. Then we split each bitstring into chunks of length $\log n/2$, perform table lookup for each chunk, and perform a prefix sum on the $O(n/\log n)$ results. We do this level-by-level so the total work for prefix sums across all levels is $O(n\log\sigma/\log n)$ and span is $O(\log n\log\sigma)$.

Setting $\tau$ to either $\sqrt{\log n\log\log n}$ or $\sqrt{\log n}$ to minimize the total work gives the following theorem:

\begin{theorem}
Wavelet matrix construction can be performed in
$O(n\log\log
n \lceil\log\sigma/ \allowbreak \sqrt{\log n\log\log n}\rceil)$ work and $O(\log n\log\sigma)$ depth
or $O((n/\epsilon)\lceil\log\sigma/\sqrt{\log n}\rceil)$ work and
$O((n^\epsilon/\epsilon)\lceil\log\sigma/\sqrt{\log n}\rceil)$
depth
for a constant $0<\epsilon<1$.
\end{theorem}

By constructing the matrix level-by-level, we can bound the working space by $O(n\log n)$ bits.

Similar to binary wavelet tree construction, we believe that a domain-decomposition approach can be used to improve the depth of the work-efficient algorithms for the variants described in this sub-section.

\section{Improved Parallel Construction of Rank/Select Structures}\label{sec:rankselect}
Wavelet trees and matrices require each node to store a succinct rank
and select structure on its bitstrings or sequences of $(\log d)$-bit values.  We
show how to construct these structures in parallel within the
bounds of the construction algorithms described in Section~\ref{sec:wt}.

\subsection{Binary Sequences}\label{sec:binary-rankselect}
We first describe the binary sequence case. The goal is to construct
the rank/select structures on $n$ bits in $O(n/\log n)$ work to match the work bound of the sequential construction algorithms in~\cite{BabenkoGKS14}. 
The overall work for rank/select construction in a wavelet
tree will therefore be $O(n\log\sigma/\log n)$, which is within the work bound
of our parallel wavelet tree algorithms. We assume that the bit sequence is packed
into $n/\log n$ words, which is provided by our wavelet tree
algorithms from Section~\ref{sec:wt}.

\myparagraph{Rank} For rank queries, we use the structure of
Jacobson~\cite{Jacobson1988}.
We only store the rank of $0$ since the rank of $1$ can be derived from the rank of the $0$.
The data structure divides the bit
sequence into \emph{ranges} of size $\log^2n$.
It computes the rank  for the last bit in
each range. The ranges are further divided into \emph{sub-ranges} of size
$\log n$, where the rank of every $\log n$'th bit relative to the
beginning of the range is stored. Inside a sub-range, the rank of a
position relative to the beginning of the sub-range can be answered
with at most two table lookups, where the table stores the answers to
all queries of sequences of up to length $\log n/2$.

We initialize an array, $A_0$, of length $n/\log n$,
and for each of the $n/\log n$ words, we count the number of $0$'s in the word and store them into its position in the appropriate
array. Counting the number of $0$'s in a word can be done in
$O(1)$ work using the same lookup table as for answering rank queries.
Then we compute the prefix sum over $A_0$. Then,
every $\log n$'th entry in $A_0$ gives the rank
for the last position in each range.  The results for the sub-ranges
are computed by taking each remaining entries in $A_0$, and
subtracting the rank stored for the closest range to the left.
The prefix sums require $O(n/\log n)$ work and
$O(\log n)$ depth.  The lookup tables can be generated in parallel in
$o(n/\log n)$ work and $O(\log n)$ depth.  The results for the
sub-ranges should be represented using $O(\log\log n)$ bits each, and
groups of $O(\log n/\log\log n)$ entries can be packed into a word as
a post-processing step in $O(n/\log n)$ work and $O(\log n)$ depth.

\myparagraph{Select} For select queries, we use Clark's select
structure~\cite{Clark1996}, which uses $O(n/\log\log n)$ extra bits for an input of length $n$.  We describe the case for querying the
location of $1$ bits, and the case for querying $0$ bits is analogous.
Clark's data structure stores the location of every $\log n\log\log
n$'th $1$ bit, which defines \emph{ranges}. For a range of length $r$
between the locations, if $r \ge \log^2n(\log\log n)^2$, then the answers to all of the
possible select queries in the range are directly stored. Otherwise, the location of
every $\log r\log\log n$'th $1$ bit is stored, which defines
\emph{sub-ranges}. For a sub-range of length $r'$, if $r' \ge
\log{r'}\log r(\log\log n)^2$ then answers are stored
directly relative to the start of the sub-range using $O(\log\log n)$ bits each. Queries that fall into all other sub-ranges are answered via
a lookup table that stores all answers for bitstrings of length $r' =
O((\log\log n)^4)$.  

To construct the select structure, we count the number of $1$'s in
each of the $2n/\log n$ half-words using table lookup, and perform a prefix
sum over the $2n/\log n$ results. We can now identify all of the half-words
that contain the location of a $k\log n\log\log n$'th $1$ bit, for any
integer $k$. Using table lookup we can find the location of the $j$'th occurrence (for a value of $j$ determined by the prefix sum) of a $1$ bit in a half-word in $O(1)$ work, which we then offset by the starting location of the half-word. This can be done in $O(n/\log n)$ work and $O(\log n)$ depth. This also allows us to determine the range lengths.
For the ranges of length at least $\log^2n(\log\log n)^2$, we scan
through the half-words in the range and store the location of every
$1$ bit. The location of all $1$ bits within a half-word can be
determined in $O(x)$ work and $O(1)$ depth via table lookup, where $x$
is the number of $1$'s in the half-word (the $O(x)$ term comes from
having to output the $x$ locations).  The locations within the
half-word are then offset by the starting location of the half-word,
again taking $O(x)$ work and $O(1)$ depth.  Scanning the half-words
takes $O(n/\log n)$ work and $O(\log n)$ depth.  There are at most
$n/(\log n\log \log n)$ locations of $1$ bits found this way, and we
can store their locations in the appropriate range in $O(n/(\log
n\log\log n))$ work and $O(1)$ depth
using the result of the previous prefix sum and subtracting the offset of where its range begins.

For ranges of length less than $\log^2n(\log\log n)^2$, we perform
a prefix sum over the half-words (as before, the count in a half-word is found via table lookup) in the range to identify
which half-words have boundaries for sub-ranges, which takes $O(n/\log n)$
work and $O(\log n)$ depth overall. Directly generating the boundary locations 
and then packing them into words would
require $O(n/(\log r\log\log n))$ work since there could be that many
locations, and this is too much. 
Instead, for the half-words that have boundaries, we
output all of the boundary locations (relative to the beginning of the range) in packed representation by using
table lookup. The lookup table takes a half-word, a skip amount $s$, an offset $j$, and a
length $r$ (these values are all bounded by the range length $\log^2n(\log\log n)^2$), and outputs the location offset by $j$ of every $s+k\log r\log\log n$'th $1$ bit for all $k$ in a packed
representation. It can be constructed by considering all possible
half-words, and all possible values of $s$, $j$, and $r$, in $o(n/\log n)$
work and $O(\log n)$ depth.  
There are at most $O(n/(\log r\log\log n))$ boundaries, and each takes $O(\log\log n)$ bits to store. We can output $O(\log n/\log\log n)$ boundaries in a word in constant work, and so outputting all of the boundaries
takes
$O(n/(\log r\log n)) = O(n/\log n)$ work and $O(\log n)$ depth.

If answers in the sub-range need to be stored directly
(i.e., the sub-range length $r'$ is at least $\log{r'}\log r(\log\log
n)^2$), then as mentioned before we store the answers relative to the start of the sub-range
using $O(\log\log n)$ bits each.
We will generate the locations of all $1$ bits relative to
the start of the range in each half-word by using table lookup, where the result is packed into groups of $O(\log
n/\log\log n)$ relative locations.
The lookup table also takes as input how much to offset each answer. The offsets can be computed via a
prefix sum over the counts of $1$ bits in the half-words.  
The number of locations of $1$ bits output is at most $O(n/(\log\log n)^2)$, and so the
number of
groups is at most $O(n/(\log n\log\log n))$. 
The last group in each
half-word might not be fully packed but this only increases the number of groups by a constant factor. The offsets for storing the groups for each
half-word can be pre-computed via prefix sums.  The lookup table takes
at most $\log^2n(\log\log n)^2$ possible offsets, and has $O(2^{\log
  n/2})$ entries per offset, so can be constructed in
$o(n/\log n)$ work and $O(\log n)$ depth.  The
overall work for this step is thus $O(n/\log n)$ and the
depth is $O(\log n)$.
Finally, for the sub-ranges of length $r' < \log {r'} \log r
(\log\log n)^2 = O((\log\log n)^4)$, the queries are answered via a
lookup table that can be computed in
$o(n/\log n)$ work and $O(\log n)$ depth.

For the select queries to work properly, all of the words inside each
range and sub-range except the last should be fully packed, but this can be
fixed with a post-processing step that generates an array of new
words, and computes for each old word where it should copy its
results in the new word using a prefix sum. 
In parallel, each new
word is then constructed sequentially from the corresponding old
words. There are a total of $O(n/\log n)$ words in total, so 
this takes $O(n/\log n)$ work and $O(\log n)$
depth.  

We have the following theorem for constructing rank/select structures
on binary sequences:

\begin{theorem}\label{thm:binary-rankselect}
The rank and select structures for a binary sequence of length $n$
packed into $n/\log n$ words can be constructed in $O(n/\log
n)$ work and $O(\log n)$ depth.
\end{theorem}

The prefix sums operate on inputs of size $O(n/\log n)$ and therefore take $O(n)$ bits of working space. The lookup tables used all contain $o(n/\log n)$ entries and take $o(n)$ bits of working space. Thus our algorithms use $O(n)$ bits of working space. 

\subsection{Generalized Rank and Select Structures}\label{sec:generalized}
In this section, we show how to construct rank and select structures
on sequences with alphabets $\sigma=O(\log^{1/3-\delta}n)$ for $\delta > 0$ (this
solution can also be used for binary sequences although the solution
described in Section~\ref{sec:binary-rankselect} is simpler).  For a
sequence of length $n$, Shun~\cite{Shun2015} describes how
to construct the structures for $O(n)$ work and $O(\log n)$ depth. We
show that the construction can be done in $O(n\log\sigma/\log n)$ work
and $O(\log n)$ depth.
While
a work bound of $O(n/\sqrt{\log n})$ suffices for use in the multiary
wavelet tree algorithm described in Section~\ref{sec:wt}, our goal is to match the work of the sequential
algorithms for constructing the generalized rank/select structures
of~\cite{BabenkoGKS14}. We assume the input is packed into
$n\log\sigma/\log n$ words.

\myparagraph{Rank} For the rank structure, a query $\rank_{\le c}(S,i)$
returns the number of times a symbol \emph{less than or equal to} $c$
appears from positions $0$ to $i$, which differs from the binary
case. Thus, simply creating $\sigma$ copies of the binary rank
structure, one for each character, will not suffice.
We will instead use the generalized rank
structure described in~\cite{BabenkoGKS14}.\footnote{Specificially, this is described in Lemma 2.3 of the conference version of~\cite{BabenkoGKS14}.}

For every $\sigma\log^2n$'th symbol
in the sequence,
the generalized rank structure of~\cite{BabenkoGKS14} stores the
$\sigma$ ranks of that symbol (there is one rank per character in the alphabet). These symbols define ranges in the sequence, and we
will refer to them as \emph{range symbols}. For each range, the
$\sigma$ ranks of every $\log n/(3\log\sigma)$'th symbol relative to
the beginning of the range are stored. These symbols define
sub-ranges, which we refer to as \emph{sub-range symbols}. Queries
inside a sub-range are of length at most $\log n/(3\log\sigma)$ and
can be answered in $O(1)$ work via table lookup. The table has
$O(\sigma^{\log n/(3\log\sigma)}\log n/(3\log\sigma))$ entries per
character, which sums to $o(n\log\sigma/\log n)$ overall, and thus can be
constructed in $o(n\log\sigma/\log n)$ work and $O(\log n)$ depth using
similar ideas as before.

We first describe how to compute the ranks of all sub-range
symbols relative to the beginning of its range.  
The algorithm requires pre-computing two lookup tables. The first table takes as input a block of $\log n/(3\log\sigma)$ symbols and outputs the generalized ranks for the last symbol in the block relative to the beginning of the block in $O(1)$ work. The second table takes as input two sets of generalized ranks relative to the beginning of the range and outputs the sum of the generalized ranks in $O(1)$ work. 
Both tables 
can be constructed in $o(n\log\sigma/\log n)$ work and $O(\log n)$ depth.  
The algorithm first passes the $\log n/(3\log\sigma)$ symbols closest to the left of (and including) each sub-range symbol to the first table. The generalized ranks relative to the beginning of the range 
can now be computed in parallel using a prefix
sum where the combining operator $\oplus$ is defined by the second lookup
table. 
Note that the combining operation is associative, as required by prefix
sum.
Over all ranges, there
are $3n\log\sigma/\log n$ symbols that we compute ranks for, and so the
prefix sum takes $O(n\log\sigma/\log n)$ work and $O(\log n)$
depth. The results can be packed tightly into words using similar ideas as before. 

To compute the generalized ranks for the range symbols, 
we first obtain the generalized ranks of the last symbol of each range relative to the beginning of the range. This can be obtained by summing the generalized ranks of the last sub-range symbol in the range with the ranks of the remaining symbols after it (relative to the last sub-range symbol) using the two lookup tables defined above.
We then 
perform a
prefix sum over these values to obtain the generalized ranks relative to the beginning of the sequence.
When combining two entries, we can simply scan
through all $\sigma$ characters (in parallel) and update their
generalized ranks. Each combining operation takes $O(\sigma)$ work and
$O(1)$ depth, and there are $O(n/(\sigma\log^2n))$ entries, giving a
total complexity of $O(n/\log^2n)$ work and $O(\log n)$ depth.  
The generalized ranks for the range symbols can now be computed by looking at the ranks of the last symbol in the previous range and updating it with the value of the range symbol.
The
overall complexity for constructing the rank structure is
$O(n\log\sigma/\log n)$ work and $O(\log n)$ depth.

\myparagraph{Select} For the select structure, we could simply create
$\sigma$ copies of the binary select structure in
Section~\ref{sec:binary-rankselect}, one per character. However, the
binary select structure that we use takes $O(n/\log\log n)$ bits of space,
and so this will not be a succinct representation for large
$\sigma$. We will therefore parallelize the construction of the
generalized select structure described in~\cite{BabenkoGKS14}. It has
been described how to do this in $O(n)$ work in~\cite{Shun2015}, but
to do this in $O(n\log\sigma/\log n)$ work to match the bound
in~\cite{BabenkoGKS14} requires additional care.

We will have a separate select structure for each
character but the structure is not the same as in the binary case. For
a character $c$, the structure stores the location of every
$\sigma\log^2n$'th occurrence of $c$, and these occurrences define
ranges (call these occurrences \emph{range symbols}). For each range,
if the length is at least $\sigma^2\log^4n$ then we store the answers
directly, and otherwise we store the locations for every
$\sigma(\log\log n)^2$'th occurrence of $c$ relative to the start of
the range, which define sub-ranges (call these occurrences
\emph{sub-range symbols}). For a sub-range, if the length is at least
$\sigma^3(\log\log n)^4$, the answers are stored directly, and
otherwise a lookup table is used to answer any query in the sub-range
in $O(1)$ work. The table contains $O(2^{\sigma^3(\log\log
  n)^4}\sigma^3(\log\log n)^4) = o(n\log\sigma/(\sigma\log n))$ entries since $\sigma
= O(\log^{1/3-\delta} n)$ for $\delta > 0$. Thus it can be constructed within the desired complexity
bounds. 

We will construct the select structures for all characters together.
We first split the input sequence into \emph{chunks} of $\log^2 n/(3\log\sigma)$
symbols and compute the number of occurrences of each character inside a chunk.
Each chunk is further split into \emph{groups} of $\log n/(3\log\sigma)$ symbols each.
We can
output the number of occurrences of each character in a group using table lookup in $O(1)$ work. 
The table contains $O(2^{\log n/3})$ entries, and thus can be computed
in $o(n\log\sigma/\log n)$ work and $O(\log n)$ depth.  
We can also use table lookup to add two sets of $\sigma$ counts together in $O(1)$ work.
Each count has a maximum value of $\log^2n/(3\log\sigma)$ and thus any count requires $O(\log\log n)$ bits to represent. The number of possible inputs to this table is therefore $2^{O(2\sigma\log\log n)}=o(n\log\sigma/\log n)$ and so the table can be constructed within the desired bounds.
To compute the number of occurrences of each character inside a chunk, we sum together the occurrences across the groups sequentially. This
takes $O(\log n)$
depth since there are $\log n$ groups per chunk.
The computation is parallelized across all chunks and the overall work performed is $O(n\log\sigma/\log n)$ and overall depth is $O(\log n)$.

Now we must find the range symbols. We perform a prefix sum over the
answers computed above, where the associative combining operator is
defined by a lookup table that takes the $\sigma$ counts from two
chunks and outputs the $\sigma$ counts that correspond to the sum of the
counts from the two input chunks. The counts here will be relative to
the beginning of the sequence, and thus an output can take $O(\sigma\log
n)$ bits to represent and $O(\sigma)$ work to output. There are
$O(n\log\sigma/\log^2n)$ chunks, and thus the prefix sum takes
$O(n\sigma\log\sigma/\log^2n) = O(n\log\sigma/\log n)$ work and
$O(\log n)$ depth.

We now know the number of occurrences of each character
in each chunk as well as from the beginning of the sequence up to that
chunk.
This allows us to identify which chunks the range symbols occur in for a given character,
and we search in the associated groups in the chunk for the
location of the range symbol. For each chunk, we scan over the
groups sequentially updating the number of times we have seen a
symbol so far via table lookup. Whenever we find a group that
contains a range symbol, we use table lookup find the location of the
$j$'th occurrence of a character inside the group in $O(1)$ work for
an appropriate value of $j$.  Thus, processing each chunk takes
$O(\log n)$ work and depth.
The lookup table can be constructed in $o(n\log\sigma/\log n)$ work
and $O(\log n)$ depth.  This process gives all of the range symbols
for a single character.  There are at most $n/(\sigma\log^2n)$ chunks
that need to be checked per character, each one taking $O(\log n)$
work. Summed across all characters, the work is $O(n/\log n)$ and the
depth is
$O(\log n)$ (we can do this process for all characters and all chunks in parallel).

With this information, we can compute the lengths of the ranges
between range symbols. For a given character $c$, for the ranges that are at least
$\sigma^2\log^4n$ long, we store all of the locations of $c$.
Finding these locations requires scanning the
relevant chunks, which takes $O(n\log\sigma/\log n)$ work and $O(\log
n)$ depth (each chunk is scanned sequentially). If we mark
the relevant chunks for each character beforehand, one scan over all
of the chunks suffices to obtain the information for all characters. 
In particular on each chunk, for each character, we mark the start and the end of the chunk that it should consider (with a special value if a character's long ranges do not span the chunk).
This information on each chunk requires $O(\sigma\log\log n) = o(\log n)$ bits and thus can be packed into a word and accessed in constant work. 
The scan over all chunks takes $O(n\log\sigma/\log n)$ work and $O(\log n)$ depth, and for each chunk we use a lookup table to find the locations of the relevant characters in each group.
The lookup table takes as input a group as well as the information stored on the chunk, and outputs the locations of all of the relevant characters
relative to the start of the group (each location is tagged with the corresponding character). The work of the query is proportional to the number of locations returned. The table has $2^{O(\sigma\log\log n)}\cdot O(2^{\log n/3})$ entries and can be constructed in $o(n\log\sigma/\log n)$ work and $O(\log n)$ depth. 
The number of locations returned is
at most $n/(\sigma^2\log^4n) \cdot \sigma\log^2n =
O(n/(\sigma\log^2n))$ per character. Summed over all characters gives
$O(n/\log^2n)$ work for returning the answers to the queries. These locations are offset by the start of the associated group.
Overall, this step takes $O(n\log\sigma/\log n)$ work and $O(\log n)$ depth.
To store these locations, we can pre-allocate space for these long ranges and compute offsets using prefix sums within the desired work and depth bounds.

For ranges of length less than $\sigma^2\log^4n$, we compute the
sub-range symbols. This process is mostly similar to how the range
symbols were computed but since there can be up to $n/(\sigma(\log\log
n)^2)$ sub-range symbols per character, outputting their locations
directly would take too much work. However, the locations only require $O(\log\log n)$ bits each so we can output $O(\log n/\log\log n)$ locations in a packed representation in constant work. We store on each chunk the start and the end of the chunk each character should consider for its short range. The lookup table takes as input a group, the information on the chunk (let $C$ be the set of characters to consider), 
a skip amount $s_c$ for each $c\in C$, and an offset $j$, and
outputs the locations offset by $j$ of every $s_c+k\sigma(\log\log n)^2$'th
occurrence of $c \in C$ in a group for all integers $k$. The offset $j$ is used to make the locations relative to the beginning of the sub-range. Both $s_c$ and $j$ are bounded by the range length, which is $\sigma^2\log^4n$. The output locations are tagged with
the corresponding character and given in a packed representation
($O(\log n/\log\log n)$ locations per word). The total work for
writing out the locations of sub-range symbols will then be
$O(n/(\sigma(\log\log n)^2)) \cdot \sigma \cdot \log\log n/\log n =
o(n\log\sigma/\log n)$.  
The lookup table can be constructed in $o(n\log\sigma/\log n)$ work and $O(\log n)$ depth. The overall work is $O(n\log\sigma/\log n)$ and depth
is $O(\log n)$.

To determine sub-ranges of length at least $\sigma^3(\log\log n)^4$,
we first return all sub-range starting locations for all characters
that satisfy this inside a group using a
lookup table.  The table will return the (packed) locations of the
sub-ranges that satisfy this property. The information on each chunk with the start and the end of the chunk in a short range for each character is also passed to the lookup table. The table can then determine which subset of characters to output sub-range starting locations for inside a group.
The number of entries in the
table is $2^{O(\sigma\log\log n)}\cdot O(2^{\log n/3})$ and can be constructed within our
complexity bounds.  However, since sub-ranges can span multiple
groups, we then use a prefix sum across all groups where
the associative operator is a lookup table that combines two
groups by keeping the latest location of each relevant character in the
first group and earliest location of each relevant character in the second. It uses the information on the chunk to determine which characters are relevant, and for which part of the groups they are relevant. It
also outputs any sub-range starting locations where the difference
between the latest location in the first group and the earliest
location in the second is at least $\sigma^3(\log\log n)^4$. This
table has $2^{O(\sigma\log\log n)}\cdot O(2^{2\log n/3})$ entries and can again be constructed
within the desired bounds. 
Without accounting for the cost of outputting the locations, the prefix sum across all groups takes $O(n\log\sigma/\log
n)$ work and $O(\log n)$ depth.
For each character, there are at most $n/(\sigma^3(\log\log n)^4)$
sub-ranges requiring answers to be stored directly, each containing
$\sigma(\log\log n)^2$ locations that require $O(\log\log n)$ bits each.
By returning the locations in packed representation, the total work is $O(n/(\sigma^3(\log\log n)^4))\cdot\sigma(\log\log n)^2\cdot\sigma\cdot\log\log n/\log n = o(n\log\sigma/\log n)$. The work for outputting the intermediate results in the prefix sum is also proportional to this.

Finally, for the remaining sub-ranges we create a
lookup table that takes a group and the information on a chunk, and outputs the position (relative to the beginning of the
sub-range) of all relevant occurrences (tagged with the
character) in a packed representation.  Constructing this table can be
done within the desired bounds. 

Overall, constructing the generalized select structure takes
$O(n\log\sigma/\log n)$ work and $O(\log n)$ depth. Combined with the algorithm for constructing the generalized rank structure, we have the following
theorem:

\begin{theorem}
For a sequence of length $n$ containing characters in
$[0,\ldots,\sigma-1]$ packed into $n\log\sigma/\log n$ words, where
$\sigma=O(\log^{1/3-\delta}n)$ for $\delta > 0$, the corresponding generalized rank and select
structures can be constructed in $O(n\log\sigma/\log n)$ work and $O(\log
n)$ depth.
\end{theorem}

The algorithms use prefix sums on inputs of length $O(n\log\sigma/\log n)$ and therefore require $O(n\log\sigma)$ bits of working space.
The lookup tables all contain $o(n\log\sigma/\log n)$ entries, therefore using $o(n\log\sigma)$ bits of working space.

\section{Conclusion}
We have described parallel algorithms for wavelet tree construction
with improved work complexity. The ideas extend to constructing
wavelet trees of arbitrary shape, multiary wavelet trees, as well as wavelet matrices.  We also
showed that the rank and select structures stored on the nodes of the
wavelet tree can be constructed work-efficiently in parallel. An
open problem is obtaining a parallel wavelet tree algorithm with
$O(n\lceil\log\sigma/\sqrt{\log n}\rceil)$ work and polylogarithmic depth for any
value of $\sigma$. 
We are also interested in improving the working space bound of some of our algorithms.

\section*{Acknowledgements}
This work was supported by the Miller Institute for Basic Research in
Science at UC Berkeley, U.S. Department of Energy Early Career Award
\#DE-SC0018947, and National Science Foundation CAREER Award
\#CCF-1845763.  The author thanks the anonymous reviewers of this
paper for helpful feedback.

\bibliographystyle{abbrv}

\bibliography{ref}

\end{document}